\documentclass[12pt]{article}
\usepackage{amsmath,amssymb,amsfonts,amsthm}

\title{Number-phase entropic squeezing and nonclassical properties of a three-level atom interacting with a two-mode field: Intensity-dependent coupling, deformed Kerr medium and detuning effects}

\author{M J Faghihi$^{1,2,3}$ and M K Tavassoly$^{1,2,*}$ \\
 \footnotesize{$^1$ Atomic and Molecular Group, Faculty of Physics, Yazd University, Yazd, Iran} \\
 \footnotesize{$^2$ The Laboratory of Quantum Information Processing, Yazd University, Yazd, Iran} \\
 \footnotesize{$^3$ Physics and Photonics Department, Graduate University of Advanced Technology, Mahan, Kerman, Iran} \\
 \footnotesize{$^*$ E-mail: mktavassoly@yazd.ac.ir}}

\begin{document}
\maketitle
 \newcommand{\norm}[1]{\left\Vert#1\right\Vert}
 \newcommand{\abs}[1]{\left\vert#1\right\vert}
 \newcommand{\set}[1]{\left\{#1\right\}}
 \newcommand{\R}{\mathbb R}
 \newcommand{\I}{\mathbb{I}}
 \newcommand{\C}{\mathbb C}
 \newcommand{\eps}{\varepsilon}
 \newcommand{\To}{\longrightarrow}
 \newcommand{\BX}{\mathbf{B}(X)}
 \newcommand{\HH}{\mathfrak{H}}
 \newcommand{\A}{\mathcal{A}}
 \newcommand{\D}{\mathcal{D}}
 \newcommand{\N}{\mathcal{N}}
 \newcommand{\x}{\mathcal{x}}
 \newcommand{\p}{\mathcal{p}}
 \newcommand{\la}{\lambda}
 \newcommand{\af}{a^{ }_F}
 \newcommand{\afd}{a^\dag_F}
 \newcommand{\afy}{a^{ }_{F^{-1}}}
 \newcommand{\afdy}{a^\dag_{F^{-1}}}
 \newcommand{\fn}{\phi^{ }_n}
 \newcommand{\HD}{\hat{\mathcal{H}}}
 \newcommand{\HDD}{\mathcal{H}}

 \begin{abstract}
 In this paper, we follow our presented model in J. Opt. Soc. Am. B {\bf 30}, 1109--1117 (2013), in which the interaction between a $\Lambda$-type three-level atom and a quantized two-mode radiation field in a cavity in the presence of nonlinearities is studied. After giving a brief review on the procedure of obtaining the state vector of the atom-field system, some further interesting and important physical features (which are of particular interest in the quantum optics field of research) of the whole system state, i.e.,  the number-phase entropic uncertainty relation (based on the two-mode Pegg-Barnett formalism) and some of the nonclassicality signs consist of sub-Poissonian statistics, Cauchy-Schwartz inequality and two kinds of squeezing phenomenon are investigated. During our presentation, the effects of intensity-dependent coupling, deformed Kerr medium and the detuning parameters on the depth and domain of each of the mentioned nonclassical criteria of the considered quantum system are studied, in detail. It is shown that each of the mentioned nonclassicality aspects can be obtained by appropriately choosing the related parameters.
 \end{abstract}

 \section{Introduction}\label{sec-intro}
 The Jaynes-Cummings model (JCM)  which is extensively used to perform a full quantum mechanical description of the atom-field interaction \cite{JCM1,JCM2}, gives a pattern to solve the interaction between the single-mode quantized electromagnetic field and the two-level atom in the rotating wave approximation (RWA).
 Relative facility of this model, in addition to appearing some non-trivial phenomena and remarkable properties such as Rabi oscillations of population inversion (which indicate the energy exchanges between field and matter), the collapses and revivals of these oscillations \cite{shore} etc gets the model in the core of this field of researches. Moreover, experimental observation of the collapse and subsequent revival, Rabi oscillations and a single-mode two-photon maser has been reported \cite{walther}.
 To expand and modify the JCM, many generalizations have been proposed in recent decades. Intensity-dependent (nonlinear) JCM was suggested in  \cite{suk1,suk2}  where the dependence of atom-field coupling on the light intensity is described, and then has been used by others \cite{selective1,selective2,selective3}.
 Dynamical behavior of the JCM beyond the RWA has been studied in \cite{naderi} in which the effects of the counter-rotating terms on various dynamical properties such as the atomic population inversion, photon counting statistics, quantum phase properties of the cavity field etc have been discussed. Quantum properties of a $\Lambda$-type three-level atom interacting with a single-mode field in a Kerr medium with intensity-dependent coupling and in the presence of the detuning parameters have been studied by us \cite{us}.
 The ability of the nonlinear JCM in generating a class of $SU(1,1)$  coherent states of the Gilmore-Perelomov type and also $SU(2)$ group was shown by one of us \cite{miry1}. In addition, as a results of a system in which a two-level atom interacts alternatively with a dispersive quantized cavity field and a resonant classical field, it is recently proposed a theoretical scheme from which the nonlinear elliptical states can be generated \cite{miry2}. In particular and in direct relation to the present work, more recently nonlinear interaction between a three-level atom (in a $\Lambda$ configuration) and a two-mode cavity field in the presence of a cross-Kerr medium and its deformed counterpart \cite{newhonarasa}, intensity-dependent atom-field coupling and the detuning parameters has been discussed by us \cite{usJOSA}. Briefly, we studied the effects of these latter parameters on a few physical properties of the obtained state vector of the entire quantum system. Specifically, using the von Neumann approach, we firstly obtained the time evolution of the field entropy by which the amount of entanglement between subsystems has been determined. In addition, the position-momentum entropic uncertainty relation is evaluated, from which the entropy squeezing in position/momentum component is investigated. \\
 Now, due to the fact that, nonclassical states have received considerable attention in various fields of research, such as quantum optics, quantum cryptography and quantum communication \cite{application1,application2,application3}, meanwhile in the continuation of our recent work in \cite{usJOSA} and for completing it,
 the main goal of the present paper is to investigate individually and simultaneously the effects of intensity-dependent coupling, deformed Kerr medium and detuning parameters on some physical properties, consist of {\it number-phase entropic squeezing},  {\it sub-Poissonian statistics},  {\it Cauchy-Schwartz inequality} (CSI) and two different types of  {\it squeezing} of the state vector of the  atom-field system under our consideration. As will be observed, the considered parameters in our model allow one to tune the depth and domain of the mentioned nonclassicality features (the aspects with no classical analogue) with respect to time. It is also noteworthy to declare that, as is well-known, there is a logical link between these nonclassicality features and Glauber-Sudarshan $P$-distribution function \cite{GSPfunction1,GSPfunction2,GSPfunction3}. By this we mean that the occurrence of nonclassicality via each of the above indicators leads to the non-positivity of the corresponding $P$-function. But, since deriving this function is usually a hard task (if not impossible), this paper is allocated to study the above nonclassicality indicators which are really independent of each other and of the previously considered properties in \cite{usJOSA}, too.
 To make our motivations of this presentation more clear, we give a few words on the notability of the considered physical (nonclassical) criteria. Starting with the number-phase entropic uncertainty relation,  it should be mentioned that, studying the quantum phase properties leads to the investigating the variation of the phase relation between the photons of the field. Moreover, it has been shown that for the single-mode JCM the evolution of the phase variance, as well as the phase distribution can carry certain information on the collapse-revival phenomenon of the corresponding atomic inversion \cite{dung}. Altogether, to achieve this purpose, it ought to be mentioned that, for discussing the dynamical behaviour of the phase properties, we will suitably extend the Pegg-Barnett formalism \cite{pb} to be utilizable for our considered model. Regarding CSI, it may be noted that, the violation of CSI may be generally led to the violation of Bell inequalities \cite{walls}. Also, it is shown that, the violation of CSI can be observed in a two-photon interference experiment with a light source in a pair coherent state \cite{ghosh}. Moreover, concerning the importance of the squeezing phenomenon, it is worthwhile to illustrate that two-mode squeezed state can be regarded as the output state of an ideal two-photon device \cite{caves}. In addition, it is reported that, in particular there exists a close link between sum squeezing phenomenon and sum-frequency generation (combination of two photons with different frequencies $\omega_{a}$ and $\omega_{b}$, and generation of a photon with frequency $\omega_{c}=\omega_{a}+\omega_{b}$, by the presence of a nonlinear medium and through the second-order susceptibility $\chi^{(2)}$ \cite{boyd}) that has been explored by Hillery \cite{hillery}. All above explanations which attracted a great deal of attention in recent researches motivate one to explore the considered nonclassicality features for the complicated atom-field interaction system which has been analytically solved by us \cite{usJOSA}. \\
 The paper is organized as follows. In the next section, a brief review of obtaining the state vector of the atom-field system will be given, which has been extensively explained in \cite{usJOSA}. In section \ref{phyproperties}, after investigating the number-phase entropic uncertainty relation by considering the two-mode Pegg-Barnett formalism, from which the entropy squeezing can be investigated, some of the nonclassical criteria of the obtained states such as Mandel $Q$ parameter, CSI and two different types of squeezing parameter are studied. Finally, section \ref{conclusion} contains a summary and concluding remarks.
%
 \section{A brief review on the model}
%
 Let us give a brief review on the model which has been extensively discussed by us in \cite{usJOSA}, in which a three-level atom (in a $\Lambda$ configuration) interacts with a two-mode quantized electromagnetic field which oscillates with frequencies $\Omega_{1}$ and $\Omega_{2}$ in an optical cavity surrounded by a Kerr medium with intensity-dependent coupling. According to the generalized JCM, the Hamiltonian for this system in the RWA can be written as $\hat{H} =\hat{H}_{0}+\hat{H}_{1}$ where ($\hbar=c=1$)
 \begin{eqnarray}\label{hamiltonih0}
 \hat{H}_{0}&=& \sum_{j=1}^{3} \omega_{j}\hat{\sigma}_{jj}+\sum_{j=1}^{2} \Omega_{j} \hat{a}^{\dag}_{j} \hat{a}_{j},
 \end{eqnarray}
 and
 \begin{eqnarray}\label{hamiltonih1}
 \hat{H}_{1} &=& \chi \hat{\R}_{1}^{\dag} \hat{\R}_{1} \hat{\R}_{2}^{\dag} \hat{\R}_{2}+\lambda_{1}(\hat{\A}_{1} \hat{\sigma}_{12}+\hat{\sigma}_{21}\hat{\A}_{1}^{\dag}) \nonumber \\
 &+& \lambda_{2}(\hat{\A}_{2}\hat{\sigma}_{13}+\hat{\sigma}_{31}\hat{\A}_{2}^{\dag}),
 \end{eqnarray}
 where $\hat{\sigma}_{ij}$ is the atomic lowering and raising operator between $|i\rangle$ and $|j\rangle$ defined by $\hat{\sigma}_{ij}=|i\rangle \langle j|,(i,j=1,2,3),\hat{a}_{j}$ ($\hat{a}_{j}^{\dag}$) is the bosonic annihilation (creation) operator of the field mode $j$, $\chi $ denotes the third-order susceptibility of cross-Kerr medium and the constants $\lambda_{1},\lambda_{2}$ determine the strength of the atom-field couplings. In above relation, $\hat{\R}_{j}=\hat{a}_{j}\hat{g}_{j}(\hat{n}_{j})$ and $\hat{\A}_{j}=\hat{a}_{j}\hat{f}_{j}(\hat{n}_{j})$ with $\hat{\R}_{j}^{\dag}$ and $\hat{\A}_{j}^{\dag}$ as their respective conjugate hermitians, where $\hat{n}_{j}=\hat{a}_{j}^{\dag}\hat{a}_{j}$. Note that, $g_{j}(\hat{n}_{j})$ and $f_{j}(\hat{n}_{j})$ are two generally different operator-valued functions which depend on the light intensity and specify the deformation of the Kerr medium and intensity-dependent atom-field coupling, respectively. \\
 The wave function $|\psi(t)\rangle$ corresponding to the whole system may be proposed in the form
 \begin{eqnarray}\label{say}
 |\psi(t)\rangle&=&\sum_{n_{1}=0}^{+\infty}\sum_{n_{2}=0}^{+\infty}q_{n_{1}}q_{n_{2}}\Big[ A(n_{1},n_{2},t) e^{-i\gamma_{1} t}|1,n_{1},n_{2}\rangle \nonumber \\
 \hspace{-2cm}&+& B(n_{1}+1,n_{2},t)e^{-i\gamma_{2} t}|2,n_{1}+1,n_{2}\rangle  \nonumber \\
 \hspace{-2cm}&+& C(n_{1},n_{2}+1,t)e^{-i\gamma_{3} t}|3,n_{1},n_{2}+1\rangle  \Big],
 \end{eqnarray}
 where $q_{n_{1}}$ and $q_{n_{2}}$ describe the amplitudes of the initial states of the field associated to each mode and $A$, $B$ and $C$ are the time-dependent atomic probability amplitudes which have to be evaluated. Also we have set
 \begin{eqnarray}
 \gamma_{1}&\doteq&\omega_{1}+n_{1}\Omega_{1}+n_{2}\Omega_{2}, \nonumber \\
 \gamma_{2}&\doteq&\omega_{2}+(n_{1}+1)\Omega_{1}+n_{2}\Omega_{2}, \nonumber \\
 \gamma_{3}&\doteq&\omega_{3}+n_{1}\Omega_{1}+(n_{2}+1)\Omega_{2}.
 \end{eqnarray}
 By applying the probability amplitudes method, and after some lengthy but straightforward manipulations we obtained the probability amplitudes $A$, $B$ and $C$ (which determine the explicit form of the wave function of whole system) in the following form \cite{usJOSA}:
 \begin{eqnarray}\label{abc}
 A(n_{1},n_{2},t)&=&-e^{-i\Delta_{2}t}\sum_{j=1}^{3}(\mu_{j}+V_{B})b_{j}e^{i\mu_{j}t}, \nonumber \\
 B(n_{1}+1,n_{2},t)&=&\sum_{j=1}^{3}\kappa_{1}\;b_{j} e^{i\mu_{j} t}, \nonumber \\
 C(n_{1},n_{2}+1,t)&=&\frac{e^{i(\Delta_{3}-\Delta_{2})t}}{\kappa_{2}}\sum_{j=1}^{3}\Big[(\mu_{j}+V_{B}) \nonumber \\
 &\times& (\mu_{j}+V_{A} - \Delta_{2}) - \kappa_{1}^{2}\Big]b_{j}e^{i\mu_{j}t},
 \end{eqnarray}
 where
 \begin{eqnarray}\label{vkardan}
 \mu_{j}&=&-\frac{1}{3}x_{1}+\frac{2}{3}\sqrt{x_{1}^{2}-3x_{2}}\cos\left[ \theta+\frac{2}{3}(j-1)\pi \right], \;j=1,2,3, \nonumber \\
 \theta &=& \frac{1}{3}\cos^{-1}\left[ \frac{9x_{1}x_{2}-2x_{1}^{3}-27x_{3}}{2(x_{1}^{2}-3x_{2})^{3/2}}\right],
 \end{eqnarray}
 with
 \begin{eqnarray}\label{x123}
 x_{1}&\dot{=}&V_{A}+V_{B}+V_{C}+\Delta_{3}-2\Delta_{2}, \nonumber \\
 x_{2}&\dot{=}& (V_{A}+V_{B}-\Delta_{2})(V_{C}+\Delta_{3}-\Delta_{2})+V_{B}(V_{A}-\Delta_{2})  \nonumber \\
 &-& \kappa_{1}^{2} - \kappa_{2}^{2}, \nonumber \\
 x_{3}&\dot{=}&V_{B}\left[(V_{A}-\Delta_{2})(V_{C}+\Delta_{3}-\Delta_{2})-\kappa_{2}^{2} \right]  \nonumber \\
 &-& \kappa_{1}^{2}(V_{C}+\Delta_{3}-\Delta_{2}).
 \end{eqnarray}
 Also, in the above relations we have defined
 \begin{eqnarray}\label{vdefinition1}
 \hspace{-0.5cm}V_{A} &\dot{=}&V(n_{1},n_{2}), \;\; V_{B}\dot{=}V(n_{1}+1,n_{2}), \;\; V_{C}\dot{=}V(n_{1},n_{2}+1), \nonumber \\
 \hspace{-0.5cm}\kappa_{1}&\dot{=}&\lambda_{1}\;\sqrt{n_{1}+1}\;f_{1}(n_{1}+1), \;\; \kappa_{2}\dot{=}\lambda_{2}\;\sqrt{n_{2}+1}\;f_{2}(n_{2}+1),  \nonumber \\
 \hspace{-0.5cm} \Delta_{2} &=& \omega_{2}-\omega_{1}+\Omega_{1},  \;\;\; \Delta_{3}=\omega_{3}-\omega_{1}+\Omega_{2},
 \end{eqnarray}
 with
 \begin{eqnarray}\label{vdefinition1}
 V(n_{1},n_{2}) \dot{=} \chi\; n_{1} n_{2} g_{1}^{2}(n_{1})g_{2}^{2}(n_{2}).
 \end{eqnarray}
 Finally, by preparing the atom initially in the excited state, i.e. $A(0)=1$, $B(0)=C(0)=0$ or equivalently
 \begin{eqnarray}\label{sayi}
 |\psi(0)\rangle_{\mathrm{A-F}}=|1\rangle \otimes \sum_{n_{1}=0}^{+\infty}\sum_{n_{2}=0}^{+\infty}q_{n_{1}}q_{n_{2}} |n_{1}, n_{2} \rangle,
 \end{eqnarray}
 the following relation for $b_{j}$  may be found
 \begin{eqnarray}\label{b123}
 \hspace{-1cm} b_{j}=\frac{\mu_{k}+\mu_{l}+V_{A}+V_{B}-\Delta_{2}}{\mu _{jk} \mu _{jl}}, \; j\neq k\neq l=1,2,3,
 \end{eqnarray}
 where $ \mu _{jk}=\mu_{j}-\mu_{k} $. Hence, the wave function $|\psi(t)\rangle$ as given in (\ref{say}) is exactly derived.
 It is valuable to state that the above formalism can be used for any physical system with arbitrary nonlinearity function. In this paper, we use the nonlinearity function $f(n)  = 1/\sqrt{n} = g(n)$ which has been derived by Man'ko {\it et al} \cite{manko2} from the coherent states that have been named by Sudarshan as harmonious states \cite{harmonious}. This function is a popular nonlinearity function which has been usually used in the contents of deformation of bosonic operators in quantum optics literature \cite{example-harmonious1,example-harmonious2,example-harmonious3}.
%
 \section{Investigating the physical properties of the introduced state}\label{phyproperties}
 As we pointed out in the Introduction of the paper, in this section, we intend to examine the further physical properties with emphasising on some nonclassical features of the obtained state (in previous section) which are of special interest in the field of quantum optics and quantum information processing. To achieve this purpose, we check number-phase entropic squeezing, sub-Poissonian statistics, CSI and two different squeezing criteria, in detail. We will observe, the state has potential ability to show high nonclassicality behaviour in appropriate intervals of time. In this way, it may be noted that the unfinished work in \cite{usJOSA}, in our opinion, will be truly completed.
%
 \subsection{Number-phase entropic uncertainty relation and entropy squeezing}\label{phase}
  It is well-known that the investigation of the dynamical behavior of phase distribution of the field photons is in fact an analysis of the variation of the phase relation between field photons and atoms \cite{nakano}. Phase distribution and squeezing in number/phase operator of various physical systems with known discrete spectrum $e_{n}$ have been reported by one of us in \cite{honar}. Also, the number-phase entropic uncertainty relation and the number-phase Wigner function of generalized coherent states associated with a few solvable quantum systems that have non-degenerate spectra are examined \cite{honarasaphase}. Here, we intend to investigate the {\it entropy squeezing} for considered bipartite system in terms of entropies of {\it number} and {\it phase} operators, in which the phase properties of a photon field are evaluated by using the Pegg-Barnett approach \cite{pb}. According to this method, all observables corresponding to the phase properties are defined in an $(s+1)$-dimensional space, in which they constitute the eigenvalue equations with $(s+1)$  eigenstates (orthonormal phase states). Based on the Pegg-Barnett formalism, a complete set of $(s +1)$ orthonormal phase states (of a single-mode field) is defined by
  \begin{eqnarray}\label{pb1mode}
  |\theta_{p} \rangle = \frac{1}{\sqrt{s+1}}\sum_{n=0}^{s}\exp(in\theta_{p})| n \rangle,
  \end{eqnarray}
  where $\{| n \rangle\}_{n=0}^{s}$ are the number states and $\theta_{p}$ gets the following values:
  \begin{eqnarray}\label{thetap}
  \theta_{p}=\theta_{0}+\frac{2\pi p}{s+1},\;\;\;\;p=0,1,...,s,
  \end{eqnarray}
  with $\theta_{0}$ as an arbitrary value. It is useful to state that in calculating the required expectation values, the parameter $s$ tends finally to infinity. Moreover, since we are dealing with the two-mode field in the present atom-field interaction, it is necessary to generalize the orthonormal phase states. Thus, we propose the two-mode phase state $|\theta_{p},\theta_{q} \rangle$ which should be necessarily introduced as
  \begin{eqnarray}\label{pq2mode}
  \hspace{-1cm} |\theta_{p},\theta_{q} \rangle = \frac{1}{s+1}\sum_{n=0}^{s}\sum_{m=0}^{s} \exp\left(i n \theta_{p}\right) \exp\left( i  m \theta_{q}\right) | n,m \rangle,
  \end{eqnarray}
  with
  \begin{eqnarray}\label{thetapq}
  \theta_{k}=\theta_{0}+\frac{2\pi k}{s+1},\;\;\;\;k=p,q,
  \end{eqnarray}
  and $\theta_{0}$ is an arbitrary value. In the same manner which stated in (\ref{thetap}), each of the indices, $p$ and $q$ can get the values $0,1,...,s$. By the way, we are able to define the two-mode Pegg-Barnett phase distribution function as follows
  \begin{eqnarray}\label{pthetapq}
  \mathcal{P}_{\theta}(\theta_{p},\theta_{q})=\lim_{s\rightarrow +\infty} \left( \frac{s+1}{2\pi} \right)^{2} \langle \theta_{p},\theta_{q}|\hat{\rho}_{F}|\theta_{p},\theta_{q} \rangle,
  \end{eqnarray}
  By putting the relation (\ref{pq2mode}) into (\ref{pthetapq}) and paying attention to the state vector of the whole system proposed in (\ref{say}), which all of its time-dependent coefficients have been explicitly obtained, one may obtain the phase distribution function associated with the two-mode cavity field as
   \begin{eqnarray}\label{pthetapqfinal}
  \hspace{-2cm}\mathcal{P}_{\theta}(\theta_{1},\theta_{2}) &=& \frac{1}{4 \pi^{2}} \left| \sum_{n_{1}=0}^{+\infty}\sum_{n_{2}=0}^{+\infty} q_{n_{1}} q_{n_{2}} A(n_{1},n_{2},t) \exp(- i n_{1} \theta_{1}) \exp(- i n_{2} \theta_{2}) \right|^{2} \nonumber \\
  \hspace{-2cm}&+& \frac{1}{4 \pi^{2}} \left| \sum_{n_{1}=0}^{+\infty}\sum_{n_{2}=0}^{+\infty} q_{n_{1}} q_{n_{2}} B(n_{1}+1,n_{2},t) \exp(- i n_{1} \theta_{1}) \exp(- i n_{2} \theta_{2})  \right|^{2} \nonumber \\
  \hspace{-2cm}&+& \frac{1}{4 \pi^{2}} \left| \sum_{n_{1}=0}^{+\infty}\sum_{n_{2}=0}^{+\infty} q_{n_{1}} q_{n_{2}} C(n_{1},n_{2}+1,t) \exp(- i n_{1} \theta_{1}) \exp(- i n_{2} \theta_{2})  \right|^{2}.
 \end{eqnarray}
Now, based on the Shannon's idea which is related to classical information theory and following the path of \cite{pb}, we may define the (Shannon) entropies associated with the number and phase probability distribution by the following relations:
  \begin{eqnarray}\label{shannonent}
  \hspace{-1cm}R_{n}(t) &=& - \sum_{n_{1}=0}^{+\infty} \sum_{n_{2}=0}^{+\infty}  \mathcal{P}_{n}(n_{1},n_{2}) \ln \mathcal{P}_{n}(n_{1},n_{2}), \nonumber \\
  \hspace{-1cm}R_{\theta}(t) &=& - \int \limits_{\theta_{0}}^{\theta_{0}+2\pi} d\theta_{1} \int \limits_{\theta_{0}}^{\theta_{0}+2\pi} d\theta_{2} \; \mathcal{P}_{\theta}(\theta_{1},\theta_{2}) \ln \mathcal{P}_{\theta}(\theta_{1},\theta_{2}),
  \end{eqnarray}
 where $\mathcal{P}_{n}(n_{1},n_{2})=\langle n_{1},n_{2}|\hat{\rho}_{F}|n_{1},n_{2} \rangle$ and $\hat{\rho}_{F}=\mathrm{Tr}_{\mathrm{Atom}} \left( |\psi(t) \rangle \langle \psi(t)| \right)$. The sum of the entropic uncertainty relations for the number and phase in (\ref{shannonent}) can specify the lower bound for entropy which is given by $R_{n}+R_{\theta} \geq \ln 2\pi$. Considering this inequality, we suggest two quantities as follows:
 \begin{eqnarray}\label{entsq}
 S_{n}(t)&=&\frac{1}{\sqrt{2\pi}} \exp(R_{n}(t))-1, \nonumber \\
 S_{\theta}(t)&=&\frac{1}{\sqrt{2\pi}} \exp(R_{\theta}(t))-1.
 \end{eqnarray}
 These quantities indicate that when $-1<S_{n}(t)<0$ ($-1<S_{\theta}(t)<0$), the number (phase) component of the field entropy is squeezed. It is worthwhile to declare that the negativity of $S_{n(\theta)}(t)$ and in other words, entropy squeezing is another expression of the fact that $R_{n(\theta)}(t)$ is below its minimum value.\\
 Figure \ref{quant.ph-sq} shows the time evolution of the entropy squeezing in phase against the scaled time $\tau$ for initial mean number of photons fixed at $|\alpha_{1}|^{2} = 10=|\alpha_{2}|^{2}$.  The upside plots concern with the absence of the intensity-dependent coupling, i.e. $f_{i} (n_{i}) = 1$  and the downside plots correspond to the intensity-dependent coupling regime by using the nonlinearity function $f_{i} (n_{i})=1/\sqrt{n_{i}}$.
 In figure \ref{quant.ph-sq}(a)  Kerr effect is absent  ($\chi = 0$) and the exact resonant case is assumed ($\Delta_{2} = \Delta_{3} =0$).
 Figure \ref{quant.ph-sq}(b) shows the effect of deformed Kerr medium ($\chi = 0.4 \lambda$), by considering the nonlinearity function $ g_{i}(n_{i})  = 1/\sqrt{n_{i}}$, in the absence of the detuning parameters. The influence of deformed Kerr medium in the presence of detuning parameters has been depicted in figures \ref{quant.ph-sq}(c).
 From the upside plot of figure \ref{quant.ph-sq}(a) which corresponds to the constant coupling regime, no Kerr medium and in the resonance condition, it is observed that the state of the system possesses the entropy squeezing until it becomes positive after nearly $\tau=35$. By entering the effect of intensity-dependent coupling (downside plot of figure \ref{quant.ph-sq}(a)), remarkable constant amount  of entropy squeezing ($\simeq -0.81954$) is appeared at all times. Considering both upside plots of figures \ref{quant.ph-sq}(a) and \ref{quant.ph-sq}(b) shows that, they are nearly the same, both qualitatively and quantitatively. Looking at  the downside plot of figure \ref{quant.ph-sq}(b) indicates that, deformed Kerr medium can strengthen the entropy squeezing. In this case, the exact periodicity of oscillations is clearly observed. In detail, while upside plots of figures \ref{quant.ph-sq}(a) and \ref{quant.ph-sq}(b) are nearly the same, entering simultaneously the effects of nonlinearities in downside plot of figure \ref{quant.ph-sq}(b) results in an increase in the amount of the negativity of entropy squeezing in contrast with the downside plot of figure \ref{quant.ph-sq}(a). Figure \ref{quant.ph-sq}(c) shows the influences of the deformed Kerr medium and detuning parameters. It is observed that, in the presence of the detuning parameters with $f_{i}(n_{i}) = 1$ (the upside plot of figure \ref{quant.ph-sq}(c)),  deformed Kerr medium generally increases the depth  and time interval of the negativity of the phase entropic squeezing parameter. This is  while in the intensity-dependent coupling regime,  the appearance of entropy squeezing at all times is observed with fast oscillatory behavior (the downside one). \\
 So, it may be inferred that, intensity-dependent coupling has a direct role in exhibiting the entropy squeezing in phase component at all times and strengthen this particular nonclassicality feature (see the downside plots). Also, deformed Kerr nonlinearity can improve this nonclassicality sign and detuning may enrich the negativity of this criterion (especially in the constant coupling). In addition, fast oscillatory behaviour of this quantity can be regarded as a consequence of the presence of the detuning parameters (in the intensity-dependent coupling regime).
%
 \subsection{Sub-Poissonian statistics}\label{mandelparameter}
 %
 In order to determine photon statistics of the field we use the Mandel's $Q$ parameter which is defined by \cite{mandel}:
 \begin{eqnarray}\label{mandel1}
 Q = \frac{\langle (\Delta \hat{n})^2\rangle - \langle \hat{n} \rangle}{\langle \hat{n} \rangle},
 \end{eqnarray}
 where $\langle (\Delta \hat{n})^2\rangle=\langle \hat{n}^{2} \rangle-\langle \hat{n} \rangle^{2}$ and $\hat{n}=\hat{a}^{\dag}\hat{a}$. This parameter states that whenever $-1\leq Q <0 \;(Q>0)$ the statistics is sub-Poissonian  (super-Poissonian) and $Q=0$ indicates the Poissonian statistics. By the way, the state vector of the system shows the nonclassical behavior when the photons statistics of field is sub-Poissonian. \\
 Our presented results in figure \ref{mandel} show the time evolution of Mandel parameter versus the scaled time $\tau$, in which photon statistics of the cavity field (mode $1$) has been studied. In plotting this figure, the upside (downside) plots again correspond to the case $f_{i} (n_{i}) = 1$ ($f _{i}(n_{i}) = 1/\sqrt{n_{i}}$), also all parameters have been chosen like figure \ref{quant.ph-sq}.
 Figure \ref{mandel}(a) is depicted for the case in which intensity-dependent coupling, deformed Kerr and detuning effects are all  neglected, i.e.,  $(f_i(n_i)=1, \chi = 0 = \Delta_{2} = \Delta_{3})$. It is observed from the upside plot that the Mandel parameter varies between negative and positive values with collapse and revival behaviour. By entering the intensity-dependent coupling, it is seen that this parameter oscillates in negative region for all times.  In figure \ref{mandel}(b), we have focused on the study of the effect of deformed Kerr medium on the time evolution of Mandel parameter. Comparing the upside plots of figures \ref{mandel}(a) and \ref{mandel}(b) indicates that the effect of deformed Kerr medium may enhance the negativity of Mandel parameter. Also, from the downside ones, the oscillatory behavior of the Mandel parameter is clearly observed, with increasing the amount of negativity of this quantity.
 The effect of detuning parameters in the presence of deformed Kerr medium has been shown in figure \ref{mandel}(c).
 According to the upside plots of figures \ref{mandel}(b) and \ref{mandel}(c), it is apparent that entering the detuning parameters in the presence of deformed Kerr medium reduces the average time intervals  of the negativity of Mandel parameter (nonclassicality behaviour). While paying attention to the downside ones shows that due to entering the detuning, Mandel parameter will be always negative although the minima values of this nonclassicality sign is diminished. \\
 From the numerical results depicted in figure \ref{mandel}, one can deduce that intensity-dependent coupling, as well as the deformed Kerr medium possesses a direct role in disappearing the classicality feature and so improving this nonclassicality indicator. Also, it is found that the depth of negativity of Mandel parameter in the presence of the detuning parameters may be decreased when intensity-dependent coupling is considered. At last, in the constant coupling regime, the typical collapse and revival exhibition as a nonclassical phenomenon can be observed.
  %
 \subsection{The Cauchy-Schwartz inequality}\label{CSInequality}
%
 In this section we pay attention to CSI for checking another aspect of the nonclassicality of the state vector of the system. In the light of the Agarwal exploration \cite{agarwal} on the link between $P$-function \cite{GSPfunction1,GSPfunction2,GSPfunction3} and CSI, he showed that if the quantity $I_{0}$, as defined by
 \begin{eqnarray}\label{CSinequality}
 I_{0}=\frac{\left( \langle a_{1}^{\dag 2}a_{1}^{2}\rangle \langle a_{2}^{\dag 2} a_{2}^{2} \rangle \right)^{1/2}}{|\langle a_{1}^{\dag}a_{1} a_{2}^{\dag} a_{2} \rangle|}-1.
 \end{eqnarray}
 gets negative (positive) values, the corresponding $P$-function would be negative (positive), i.e. the nonclassical (classical) behavior is appeared. Notice that, commonly, one may expect that Glauber-Sudarshan $P$-distribution function, in the sense that it is a distribution function, have to be positive, while there exists quantum states for which $P$-function is negative or highly singular (nonclassical states).\\
 In figure \ref{csi} the time evolution of the CSI against the scaled time $\tau$ is illustrated for different chosen parameters similar to figure \ref{quant.ph-sq}, where the upside (downside) plots refer to the constant (intensity-dependent) coupling regime.
 The upside plot of figure \ref{csi}(a) shows that the quantum system is not sensitive to this nonclassical criterion when the atom-field coupling is constant, while in the presence of intensity-dependent coupling (downside plot), CSI always gets negative value with oscillatory behaviour.
 In figure \ref{csi}(b) where the effect of deformed Kerr medium is examined it is seen that,  the deformed Kerr medium has no remarkable effect on the negativity of CSI, either in the presence or absence of the intensity-dependent coupling.
 In figure \ref{csi}(c) the effect of deformed Kerr medium together with the detuning parameters is shown. From the upside plot of this figure it is observed that, CSI varies between positive (classical behaviour) and negative (nonclassical behaviour) values. Also, from downside one, this quantity takes negative values in all time domain with fast oscillatory behaviour. Comparing the upside plots of figures \ref{csi}(b) and \ref{csi}(c) indicates that due to the presence of detuning, nonclassicality behaviour is revealed. \\
 By considering the presented results depicted in figure \ref{csi}, one can deduce that intensity-dependent coupling (the downside plots) plays an effective role on appearing the negativity of CSI, as the detuning. Also, deformed Kerr medium has no notable effect on the negativity of CSI. In addition, for all cases concerning with constant coupling regime (upside plots) the (typical) collapse and revival is clearly observed.
  %
 \subsection{Squeezing}\label{squeezing}
 Squeezing phenomenon, as the last nonclassicality feature which will be discussed, is one of the applied nonclassicality features which is described by decreasing the quantum fluctuation in one of the field quadratures below the value of vacuum or canonical coherent states. In the continuation, two kinds of squeezing parameters are introduced and in each case, the squeezing condition of the state vector of the whole system is studied.
%
 \subsubsection{Two-mode squeezing}
 %
 In order to study the two-mode squeezing (since we are dealing with the two-mode cavity field) of the state vector of the system, the following (Hermitian) two-mode quadrature operators have been defined \cite{knight}
 \begin{eqnarray}\label{lsp}
 \hat{X}_{1} &=& \frac{1}{2\sqrt{2}}(\hat{a}_{1}+\hat{a}^{\dag}_{1}+\hat{a}_{2}+\hat{a}^{\dag}_{2}), \nonumber \\
 \hat{X}_{2} &=& \frac{1}{2i\sqrt{2}}(\hat{a}_{1}-\hat{a}^{\dag}_{1}+\hat{a}_{2}-\hat{a}^{\dag}_{2}).
 \end{eqnarray}
 These definitions lead to the commutation relation $[\hat{X}_{1},\hat{X}_{2}]=i/2$. Consequently, uncertainty relation for such operators read as
 $\langle (\Delta \hat{X}_{1})^{2} \rangle \langle (\Delta \hat{X}_{2})^{2} \rangle \geq 1/16$, where  $\Delta \hat{X}_{1}$ and $\Delta \hat{X}_{2}$ are the uncertainties in $\hat{X}_{1}$ and $\hat{X}_{2}$, respectively. A state is squeezed in $\hat{X}_{1} (\hat{X}_{2})$ if $\langle (\Delta \hat{X}_{1})^{2} \rangle <0.25\; \left( \langle (\Delta \hat{X}_{2})^{2}\rangle <0.25\right)$, or equivalently if $S_{X_{1(2)}}=4 \langle (\Delta \hat{X}_{1(2)})^{2} \rangle  -1$ satisfies the inequality $-1<S_{X_{1(2)}}<0$. The latter inequality implies that in order to exist squeezing, two-mode Glauber-Sudarshan $P$-function, $P(\alpha,\beta)= P(\alpha) P(\beta)$, requires to be nonpositive or singular in some regions of phase space \cite{knight}.
 Anyway, the two-mode squeezing parameters can be written as:
 \begin{eqnarray}\label{vsqdx2}
 S_{X_{1}}&=&\mathrm{Re} \Big( \langle \hat{a}_{1}^{2} \rangle + \langle \hat{a}_{2}^{2} \rangle
 +2 \langle \hat{a}_{1}^{\dag}\hat{a}_{2}\rangle+2\langle \hat{a}_{1}\hat{a}_{2} \rangle \Big)
 +\langle \hat{a}_{1}^{\dag}\hat{a}_{1} \rangle  \nonumber\\
 &+& \langle \hat{a}_{2}^{\dag}\hat{a}_{2} \rangle - 2\Big[\mathrm{Re}\Big(\langle \hat{a}_{1} \rangle+\langle \hat{a}_{2} \rangle\Big)\Big]^{2}, \nonumber\\
 S_{X_{2}}&=&\mathrm{Re} \Big( 2\langle \hat{a}_{1}^{\dag}\hat{a}_{2} \rangle
 -2\langle \hat{a}_{1}\hat{a}_{2} \rangle
 -\langle \hat{a}_{1}^{2} \rangle - \langle \hat{a}_{2}^{2} \rangle\Big)
 +\langle \hat{a}_{1}^{\dag}\hat{a}_{1} \rangle  \nonumber \\
 &+& \langle \hat{a}_{2}^{\dag}\hat{a}_{2} \rangle - 2\Big[\mathrm{Im}\Big(\langle \hat{a}_{1} \rangle+\langle \hat{a}_{2} \rangle\Big)\Big]^{2}.
 \end{eqnarray}
 Figure \ref{two-modesq} shows the temporal behavior of two-mode squeezing in $X_{1}$  for some of the chosen parameters as considered in figure \ref{quant.ph-sq}. Also, the upside (downside) plots concern with the absence (presence) of the intensity-dependent coupling. In the upside plot of figure \ref{two-modesq}(a), where the Kerr effect and intensity-dependent coupling are both absent and the resonance case is considered, the state of the system does not exhibit two-mode squeezing except in some small intervals of time. However, entering the intensity-dependent coupling leads to the two-mode squeezing  in all times (the downside one).
 The upside plots of figures \ref{two-modesq}(b) (studying the effect of deformed Kerr medium) and \ref{two-modesq}(c) (exhibiting simultaneously the effects of deformed Kerr medium and detuning parameters) show a temporal behaviour similar to the upside plot of figure \ref{two-modesq}(a).
 From the downside plot of figure \ref{two-modesq}(b), it is observed that the presence of deformed Kerr medium and intensity-dependent coupling effects brings about to appear deeply two-mode squeezing. In this case, it is valuable to state that the amount of the negativity of two-mode squeezing for this situation becomes nearly $20$ times greater than the downside plot of figure \ref{two-modesq}(a).
  From the downside plot of figure \ref{two-modesq}(c),  it seems that in the intensity-dependent coupling regime, adding the detuning parameters on the deformed Kerr medium reduces the amount of the negativity of this nonclassicality sign. \\
 Finally, it is deduced that intensity-dependent coupling helps clearly to occur the two-mode squeezing in $X_{1}$ and particularly the deformed Kerr medium strengthens this nonclassicality feature. Also, the reduction in the negativity $S_{X_{1}}$ and fast oscillatory behaviour are the results of considering the detuning effect.
  %
 \subsubsection{Sum squeezing}
 Here, we are going to investigate sum squeezing phenomenon. In order to study this nonclassicality indicator, two Hermitian operators $\hat{Y}_{1}$ and $\hat{Y}_{2}$ are given by \cite{hillery}
 \begin{eqnarray}\label{ss1}
 \hat{Y}_{1} = \frac{1}{2}(\hat{a}_{1}\hat{a}_{2}+\hat{a}^{\dag}_{1}\hat{a}^{\dag}_{2}), \;
 \hat{Y}_{2} = \frac{1}{2i}(\hat{a}_{1}\hat{a}_{2}-\hat{a}^{\dag}_{1}\hat{a}^{\dag}_{2}),
 \end{eqnarray}
 where $[\hat{Y}_{1},\hat{Y}_{2}]=\frac{i}{2}(\hat{a}^{\dag}_{1}\hat{a}_{1}+\hat{a}^{\dag}_{2}\hat{a}_{2}+1)$ yields the uncertainly relation $\langle (\Delta \hat{Y}_{1})^{2} \rangle \langle (\Delta \hat{Y}_{2})^{2} \rangle \geq\frac{1}{16}|\langle \hat{a}^{\dag}_{1}\hat{a}_{1}+\hat{a}^{\dag}_{2}\hat{a}_{2}+1\rangle)|^{2}$.
 Similar to the previous section, we can define the normalized sum squeezing parameters correspond to operators $Y_{1}$ and $Y_{2}$ respectively as
 \begin{eqnarray}\label{ssqy2}
 \hspace{-1cm}S_{Y_{1}}&=&\frac{2 \mathrm{Re} \langle \hat{a}_{1}^{2} \hat{a}_{2}^{2} \rangle+2 \langle \hat{a}_{1}^{\dag}\hat{a}_{1}\hat{a}_{2}^{\dag}\hat{a}_{2}  \rangle -4 \left( \mathrm{Re} \langle \hat{a}_{1} \hat{a}_{2} \rangle \right)^{2} }{\langle \hat{a}_{1}^{\dag}\hat{a}_{1} \rangle+ \langle \hat{a}_{2}^{\dag}\hat{a}_{2} \rangle +1}, \nonumber\\
 \hspace{-1cm}S_{Y_{2}}&=&\frac{ 2 \langle \hat{a}_{1}^{\dag}\hat{a}_{1}\hat{a}_{2}^{\dag}\hat{a}_{2}  \rangle - 2 \mathrm{Re} \langle \hat{a}_{1}^{2} \hat{a}_{2}^{2} \rangle -4 \left( \mathrm{Im} \langle \hat{a}_{1} \hat{a}_{2} \rangle \right)^{2} }{\langle \hat{a}_{1}^{\dag}\hat{a}_{1} \rangle+ \langle \hat{a}_{2}^{\dag}\hat{a}_{2} \rangle +1}.
 \end{eqnarray}
 Figure \ref{sumsq} describes the sum squeezing for different chosen parameters assumed in figure \ref{quant.ph-sq} with constant (intensity-dependent) coupling relating to the upside (downside) plots. It is seen from the upside plot ($f_{i}(n_{i}) = 1$) of figure \ref{sumsq}(a) that in the absence of the Kerr medium and in the resonance condition, the state of the system does not have sum squeezing property. However, after entering the intensity-dependent coupling, sum squeezing will be appeared in all times (the downside one).
 Figure \ref{sumsq}(b) indicates the effect of the deformed Kerr medium, from which it is found that the squeezing does not occur when the atom-field coupling is constant. On the contrary, in the intensity-dependent coupling regime, sum squeezing with oscillatory behavior is clearly observed at all times. Looking quantitatively at figures \ref{sumsq}(a) and \ref{sumsq}(b) implies the fact that, deformed Kerr medium has no remarkable effect in the negativity of sum squeezing. The effects of deformed Kerr medium and detuning parameters are simultaneously studied in figure \ref{sumsq}(c). As is seen, positive values of sum squeezing indicate the classical behaviour and so the state of quantum system does not possess this nonclassicality indicator. Comparing the downside plots of figures \ref{sumsq}(b) and \ref{sumsq}(c) shows that sum squeezing disappears when the detuning is considered. \\
 At last, one can conclude that while intensity-dependent coupling plays the important role to reach this nonclassical behaviour (negativity of sum squeezing parameter), the existence of the detuning destroys the sum squeezing of the state of the system even in the presence of intensity coupling. In addition, deformed Kerr medium preserve the amount of the negativity of this nonclassicality feature.
%
 \section{Summary and conclusion}\label{conclusion}
%
  In this paper, we have studied further physical aspects of a $\Lambda$-type three-level atom interacting with a quantized two-mode radiation field in a cavity including deformed Kerr medium  in the presence of the detuning parameters with and without atom-field intensity-dependent coupling regime.
  We have demonstrated the importance and notability of the considered nonclassicality features (which are of special interest in the quantum optics field of research) in the Introduction of this paper.
  In summary, using the analytical solution of the state vector of the considered bipartite (atom-field) system which has been very recently obtained by us in \cite{usJOSA}, first, the number-phase entropic uncertainty relation and subsequently entropy squeezing (in phase component) have been evaluated, by applying the two-mode Pegg-Barnett approach. In the continuation, other nonclassical properties of the state vector of the entire system namely, Mandel parameter, Cauchy-Schwartz inequality, Two-mode and sum squeezing have been numerically examined. In each case, we individually and simultaneously studied the effects of $` `$intensity-dependent coupling$"$ (entered by $f_{i}(n_{i})$), $` `$deformed Kerr medium$"$ (by the nonlinearity function $g_{i}(n_{i})$) and $` `$detuning parameters$"$ on the nonclassicality criteria.
 Summing up, the main results of the paper are listed in what follows.
 \begin{itemize}

 \itemsep1em

 \item {\it Tuning the nonclassicality indicators:}  It is illustrated that the amount of considered nonclassicality features can be tuned by choosing the nonlinear parameters related to the atom-field system, suitably.
   \item {\it Intensity-dependent coupling:} Presented results demonstrate that intensity-dependent coupling (which is considered by the function $f_{i}(n_{i}) = 1/\sqrt{n_{i}}$) has generally a direct role and striking effect in exhibiting and ameliorating the nonclassicality features. Even though, it seems that entering the intensity-dependent coupling in calculating Mandel parameter and CSI disappears the collapses and revivals phenomena as a nonclassicality feature.
       \item {\it Deformed Kerr medium:} Paying attention to the related results implies that the deformed Kerr medium has an obvious role and noteworthy effect in appearing and improving the nonclassicality features. Also, it is worthwhile to accentuate the fact that deformed Kerr medium which is one of the new features of the presented work, is distinguished from a Kerr medium by the nonlinearity function $g_{i}(n_{i}) = 1/\sqrt{n_{i}}$.
   \item {\it Detuning parameters:} Looking deeply at the obtained results shows that the detuning parameters quantitatively reduce the amount of the nonclassicality features. Also, as a result of the presence of the detuning parameters in the intensity-dependent coupling regime, fast oscillatory in the nonclassicality signs can be observed.
 \end{itemize}
 Finally, we would like to mention that this study can be performed for any physical system, either any nonlinear oscillator algebra with arbitrary nonlinearity function or any solvable quantum system with known discrete energy spectrum $e_{n}$ \cite{en1,en2,en3,en4,en5}, too.
 \begin{flushleft}
 {\bf Acknowledgements}\\
 \end{flushleft}
 The authors would like to express their thanks sincerely to Dr Mohammad Reza Hooshmandasl for his valuable help in the numerical results.


\end{document}